\documentclass[12pt]{iopart}

\usepackage[T1]{fontenc}
\usepackage[utf8]{inputenc}
\usepackage[english]{babel}

\usepackage{xspace}
\usepackage{siunitx}
\usepackage{hyperref}
\usepackage{graphicx}

\usepackage{multirow}

\newcommand{\HeHp}{HeH$^+$\xspace}

\newcommand{\ie}{i.\,e.\xspace}
\newcommand{\eg}{e.\,g.\xspace}
\DeclareSIUnit{\au}{{a.u.}}

\usepackage{etoolbox}

\usepackage{color}
\newcommand{\manfred}[1]{{#1}}

\begin{document}
\title[Mass-ratio dependent strong-field dissociation of helium hydride isotopologues]{Mass-ratio dependent strong-field dissociation of artificial helium hydride isotopologues}
\author{F Oppermann$^1$, S Mhatre$^2$, S Gr\"afe$^{2,3}$, M Lein$^1$}
\address{$^1$ Leibniz University Hannover, Institute of Theoretical Physics, Appelstraße 2, 30167 Hannover, Germany}
\address{$^2$ Institute of Physical Chemistry, Friedrich Schiller University Jena, Helmholtzweg 4, 07743 Jena, Germany}
\address{$^3$ Fraunhofer Institute of Applied Optics and Precision Engineering, Albert-Einstein-Str. 7, 07745 Jena, Germany}
\ead{lein@itp.uni-hannover.de}
\begin{abstract}
We study the effect of the nuclear-mass ratio in a diatomic molecular ion on the dissociation dynamics in strong infrared laser pulses. A molecular ion is a charged system, in which the dipole moment depends on the reference point and therefore on the position of the nuclear center of mass, so that the laser-induced dynamics is expected to depend on the mass asymmetry. 
Whereas
usually both the reduced mass and the mass ratio are varied when different isotopologues are compared, we fix the reduced mass and artificially vary the mass ratio in a model system.
This allows us to separate effects related to changes in the resonance frequency, which is determined by the reduced mass, from those that arise due to the mass asymmetry.
Numerical solutions of the time-dependent Schr\"odinger equation are compared with classical trajectory simulations. 
We find that at a certain mass ratio, vibrational excitation is strongly suppressed, which decreases the dissociation probability by many orders of magnitude.
\end{abstract}

\vspace{2pc}
\noindent{\it Keywords\/}: strong laser fields, helium hydride molecular ion, laser-induced dissociation, time-dependent Schrödinger equation

\maketitle

\section{Introduction}

Molecules under the influence of an external field can undergo various types of excitation and, in the case of a strong laser field, can be ionized or dissociated \cite{Posthumus2004,Holmegaard2010,Palacios2015,Ibrahim2018,Wustelt2020review}. For long wavelengths of the laser field, a large number of photons is needed to overcome the electronic excitation energies and hence, electronic transitions become unlikely. The direct excitation of atomic motion within the electronic ground state, on the other hand, requires (within the electric-dipole approximation) the presence of a non-zero permanent electric dipole or at least a change of the dipole with the geometry of the molecule.
For this reason, direct vibrational excitations are dipole-forbidden in homonuclear diatomic molecules due to their strictly vanishing electric dipole. 
\manfred{In contrast, heteronuclear diatomic molecules have a dipole that couples directly to the applied field, which means that long-wavelength fields are a suitable tool to drive atomic motion in such polar systems.}
The permanent dipole moment and the nuclear masses are the relevant parameters that determine the quantitative amount of vibrational excitation and dissociation. 

\manfred{
Applying intense laser fields to molecules is of particular interest because the high intensity opens up multiphoton pathways and provides the possibility to control chemical reactions by interference of pathways \cite{Brumer1986}. At the same time, strong fields have the ability to ionize the target, producing molecular ions in the presence of the laser field, potentially giving rise to laser-induced dissocation. The investigation and control of dissociative ionization has been an important area of research for many years \cite{Codling1993,Kumar1994,Itakura2003,Wells2009,Koh2020,Endo2022,Hasegawa2022}. However, the dynamics of polar molecular ions exhibits an important aspect that does not seem to have received attention.} For neutral diatomic molecules, we are used to the idea that the permanent dipole at a specified internuclear vector has a well-defined value independent of the nuclear masses. For a molecular ion, \manfred{on the other hand,} we must take into account the fundamental statement that the dipole moment of a charged system depends on the choice of reference point. According to the separation into center-of-mass motion and relative motion, the relevant dipoles for vibrational excitation (i.e., excitation of the relative motion) must be calculated with the center of nuclear mass as the reference point. Thus, the dipole moments of molecular ions depend on the nuclear masses. In a diatomic system, the reduced nuclear mass is, besides the dipole, the other crucial parameter that determines the nuclear dynamics. The reduced mass determines the vibrational energy levels and therefore the values of the resonant transition frequencies. \manfred{It is} important to note that the dipole moment depends on the nuclear masses even when the reduced mass is kept constant.

Isotope effects in photodissociation processes have always been a matter of interest, and many different molecular species have been studied, ranging from simple diatomic molecules \cite{Tsong1985,Jiang2019} to polyatomic or\-gan\-ic molecules \cite{Bespechansky2006,Ekanayake2018}. Nevertheless, the effect of the mass-dependent electric dipoles has not been isolated since one requires a charged system, and furthermore, \manfred{disentangling this effect} from changes in the reduced mass is not straightforward. 
\manfred{To demonstrate the impact of mass-dependent dipoles on the vibrational dynamics in its purest form, one would like to keep the reduced mass fixed (thus keeping also the vibrational level structure fixed), while varying the ratio of the nuclear masses (and thus varying the dipole moment). Therefore
it would be interesting to observe the  isotopologue dependence of laser-induced dissociation of a molecular ion at fixed reduced mass.} Unfortunately, there are few realistic target systems for such a purpose. \manfred{One may consider two isotopologues of the carbon monoxide ion, namely $^{12}$C$^{18}$O$^+$ and $^{13}$C$^{16}$O$^+$, that have almost equal reduced masses. They might serve as a set of example systems for future investigations.}  In the present work, we consider artificial isotopologues of the \HeHp molecular ion, which can be considered the simplest heteronuclear molecule. \HeHp can be prepared in the laboratory \cite{HognessLunn1925} and it has recently been observed in interstellar space \cite{Guesten2019}. It has already served as an asymmetric polar benchmark system in a number of studies \cite{Bergson1997,Saenz2003,Dumitriu2009,PhysRevLett.127.043202}. 
Its isotopologues, \eg $^4$HeH$^+$, $^4$HeD$^+$, $^3$HeH$^+$, etc. possess the same electronic configuration but they differ in both total and reduced nuclear mass. 
Most of the isotopologues that could in principle be constructed from the real isotopes of He and H differ in reduced mass, but there are two examples, $^3$HeT$^+$ and $^6$HeD$^+$, with (approximately) the same reduced mass---albeit not easily experimentally available. ($^6$He has a half-life time of \SI{0.8}{s} \cite{Audi2017}.)
In Table \ref{tab:isotopologues}, we show the properties of selected isotopologues of \HeHp, where, for simplicity, protons and neutrons are idealized as having equal masses $m_n = \SI{1837}{\au}$ and the binding energy (mass defect) of the nuclei is neglected. \manfred{We have verified that this idealization has negligible effect on the observables (dissociation probabilities) calculated in this article.} The reduced mass is given by $\mu = m_\mathrm{H} m_\mathrm{He}/M$ and we define the mass ratio $r$ as $r = m_\mathrm{H} / M$, where \(M = m_\mathrm{H} + m_\mathrm{He}\) is the total nuclear mass. \manfred{For the molecular ion $^4$He$^{1\!}$H$^+$, for example, we have $r=0.2$ in our idealization and this is very close to the exact mass ratio $r=0.2011$.}

\begin{figure}[b]
    \centering
    \includegraphics{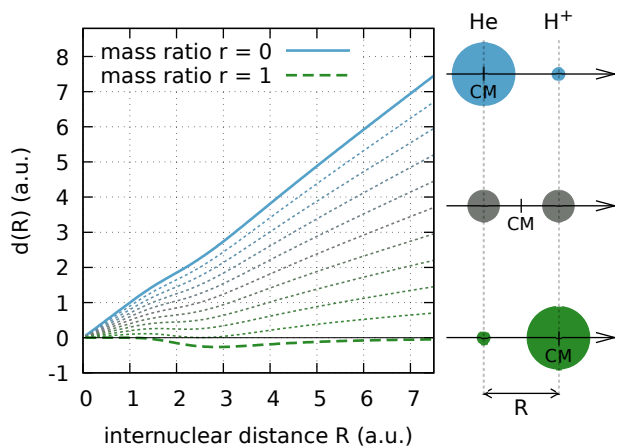}
    \caption{Left: Dipole coupling $d(R)$ in the electronic ground state for several values of the mass ratio $r$ between 0 and 1 in steps of \num{0.1}.
    Right: Changing the mass ratio moves the nuclear center of mass (CM) from the helium nucleus to the proton. The masses of the nuclei are indicated by the size of the circles. Note that the total mass diverges for \(r \to 0\) or \(r \to 1\).}
    \label{fig:dipole_coupling}
\end{figure}

In the present work, we study the two cases of $\mu = 0.8 m_n$ and $\mu = 1.5 m_n$.
We vary the mass ratio from 0 to 1, meaning that the nuclear center of mass moves from the helium nucleus to the hydrogen nucleus, see the illustration in \fref{fig:dipole_coupling}.
For fixed reduced mass, both extreme values of $r$ correspond to infinite total mass, \manfred{but the behaviour of such a system is still molecule-like since the vibrational energy levels remain the same and the dipole moment is finite.}
Using a one-dimensional model of \HeHp, we investigate the laser-induced dissociation as a function of the mass ratio. Our central result is that a strong suppression of dissociation is found for values of the mass ratio where the electric dipoles are small. We analyze this effect using both quantum-mechanical and classical simulations. 

\Table{\label{tab:isotopologues}Properties of selected isotopologues of \HeHp.}
\begin{tabular}{c c c c}\br
    isotopologue & total mass & reduced mass & mass ratio\\
		& & & $r = m_\mathrm{H} / M$\\ \mr
    $^4$HeH$^+$ & $5 m_n$ & $0.8 m_n$ & 0.2 \\ \hline
    $^4$HeD$^+$ & $6 m_n$ & $1.33 m_n$ & 0.33 \\ \hline
    $^3$HeT$^+$ & $6 m_n$ & \multirow{2}{*}{$1.5 m_n$} & 0.5 \\
    $^6$HeD$^+$ & $8 m_n$ && 0.25 \\\br
\end{tabular}
\endTable

\section{Methods}

\subsection{Electron-nuclear non-Born-Oppenheimer time-dependent Schr\"odinger equation}
\label{nonbotdse}
We apply a one-dimensional single-active-electron non-Born-Oppenheimer model and solve the time-dependent Schrödinger equation (TDSE) \cite{PhysRevLett.121.073203,Oppermann2020}.
This model covers two degrees of freedom: electronic motion along the molecular axis described by the electron coordinate $x$ (electron position relative to the nuclear center of mass) and the nuclear motion described by the internuclear distance $R$.
A softcore potential for the electron-nuclear interaction is chosen such that for frozen nuclei, the two lowest potential-energy curves match the literature values \cite{Green_1974,Pachucki2012}.
Previously, this model has been applied to various problems, including comparisons with experimental data and the control of dissociation and ionization with two-color fields \cite{PhysRevLett.121.073203,Oppermann2020}.
The wave function is represented on a grid with 2048 grid points spaced by \SI{0.05}{\au} along the $R$-axis and 4096 grid points spaced by \SI{0.2}{\au} along the $x$-axis.
    The time step for the propagation is \SI{0.02}{\au} The wave function is propagated using the split-operator method~\cite{FEIT1982412} and the
    initial states before interaction with the external field are calculated as eigenstates of the real-time evolution operator \cite{Oppermann2022}. The real-time evolution starts from the ground state. The laser pulse is modelled by defining a vector potential $A(t)$ with a $\cos^2$ envelope,
    \begin{equation}
        A(t) = \frac{E_0}{\omega}\, \cos^2(\pi t / T)\, \sin(\omega t),
    \end{equation}
    where \(T = T_{\mathrm{FWHM}}/0.3641\) and \(T_{\mathrm{FWHM}}\) is the full width at half maximum of the intensity, chosen as \ensuremath{T_{\mathrm{FWHM}}\approx\SI{50}{\femto\second}}. The pulse is linearly polarized along the molecular axis.
    The vector potential determines the electric field as $E(t) = - \dot A(t)$.
    The dissociation yield is calculated from the wave function at the end of the time evolution by first projecting out all bound states followed by projection onto electronic eigenstates 
    so that dissociation into different electronic channels can be distinguished.
\subsection{Born-Oppenheimer TDSE}
While the non-Born-Oppenheimer model allows us to describe arbitrary electronic excitations and even ionization, often this is not needed.
For comparison and as a simpler model, we apply the Born-Oppenheimer approximation and solve the TDSE only for the nuclear wave functions on two
coupled potential curves.
Although the coupling between two electronic states is included in the quantum-mechanical simulations, for which we present results below, we begin by writing the TDSE for the situation when this coupling is neglected. In this case, the TDSE for the nuclear wave function \(\psi_k(R)\) on the \(k\)-th potential-energy curve \(V_k\) reads
    \begin{eqnarray}
        \rmi \frac{\partial}{\partial t} \psi_k(R;t) = H_k(t) \psi_k(R;t), \label{eq:TDSE_1D} \\
        H_k(t) = \frac{P^2}{2 \mu} + V_k(R) - d_k(R) E(t).\label{eq:nuclear_Hamiltonian} 
    \end{eqnarray}
    Here and in the following, atomic units are used if not stated otherwise. The dipole moments \(d_k(R)\) are calculated from the model outlined in section \ref{nonbotdse}.
    To this end, the \(k\)-th electronic eigenstate \(\phi_k(x;R)\) is calculated for frozen nuclei at the internuclear distance \(R\). Since the electron coordinate $x$ is defined relative to the nuclear center of mass, the functions \(\phi_k(x;R)\) depend on the mass ratio $r$ via a coordinate shift,
    \begin{equation}
        \phi_k(x;R) = \phi_{k,r=0}\,(x+rR;R).
    \end{equation}
    Therefore, the purely electronic dipole transition moments, defined as
    \begin{equation}
        d_{jk}(R) = -\langle \phi_j \mid  x \mid \phi_k\rangle_{(x)},
    \end{equation}
    satisfy
    \begin{equation}
        d_{jk}(R) = d_{jk}(R)\Big|_{r=0} + rR\, \delta_{jk}.
        \label{eq:dipole_matrix_element}
    \end{equation}
    For the total dipole moments needed in the Born-Oppenheimer Hamiltonian (\ref{eq:nuclear_Hamiltonian}), both the electron dipole and the charged cores must be taken into account \cite{PhysRevLett.121.073203,Oppermann2020},
    \begin{equation}
        d_k(R) = - \langle \phi_k \mid (\kappa x + \lambda R) \mid \phi_k\rangle_{(x)},
    \end{equation}
    where \(\kappa = (M + 2)/(M + 1)\) and \(\lambda = (m_\mathrm{H} - m_\mathrm{He})/M = 2r - 1\). 
    Hence the dependence on $r$ can be written explicitly as
    \begin{eqnarray}
        d_k(R)  = \kappa d_{kk}(R)\Big|_{r=0} + [1-(2-\kappa)r] \,R.  \label{eq:dipole_coupling}
    \end{eqnarray}
    The (permanent) ground-state dipole moment $d_1(R)$ is simply called $d(R)$ in the following. This function is shown in \fref{fig:dipole_coupling} for various choices of the mass ratio $r$. As motivated above, the dipole moment depends strongly on the mass ratio $r$.
    
    In the two-level Born-Oppenheimer calculations, the light-induced coupling of the lowest electronic states is included in the TDSE, which reads
    \begin{eqnarray}
        \rmi \frac{\partial}{\partial t} \left(\hspace{-0.5em}\begin{array}{c}\psi_1(R;t)\\\psi_2(R;t)\end{array}\hspace{-0.5em}\right)%
            = \left(\hspace{-0.5em}\begin{array}{cc}H_1(t) & -\kappa d_{12}E(t)\\ -\kappa d_{12}E(t) & H_2(t)\end{array}\hspace{-0.5em}\right)
            \left(\hspace{-0.5em}\begin{array}{c}\psi_1(R;t)\\\psi_2(R;t)\end{array}\hspace{-0.5em}\right) \label{eq:TDSE_two-level}
    \end{eqnarray}
    with \(H_1\), \(H_2\) given by \eref{eq:nuclear_Hamiltonian} and \(d_{12}(R)\) defined in \eref{eq:dipole_matrix_element}. 
    Equation (\ref{eq:TDSE_two-level}) is solved by applying the split-operator scheme on $R$-grids with 2048 grid points spaced by \SI{0.05}{\au}, combined with the matrix exponential for the offdiagonal part of the Hamiltonian matrix. The time evolution starts from the vibrational ground state of the lowest electronic state.
    At the end of the time evolution, all bound states are projected out from
    \(\psi_1\) and the norm squared of the remaining wave function is the probability for dissociation into the electronic ground-state.
    The squared norm of \(\psi_2\) is the probability for dissociation into the first excited state.

\subsection{Classical Calculations}
    Classical trajectory Monte Carlo (CTMC) simulations are done to investigate the classical anologue of the previously described quantum system. Similar to the Born-Oppenheimer TDSE simulations, the system is defined as a particle on a Born-Oppenheimer potential and the classical Hamiltonian reads the same as in \eref{eq:nuclear_Hamiltonian}. 
   Here, we consider only a one-level system, \ie the system is assumed to stay in the electronic ground state. 
   The time evolution involves solving Newton's equations of motion,
   \begin{eqnarray}{\label{Newton's eqn}}
    \frac{\mathrm{d}P}{\mathrm{d}t} = F(R,t) = - \frac{\partial}{\partial R} \,\Big(V(R) - d(R)E(t)\Big), \\
    P =  \mu\, \frac{\mathrm{d}R}{\mathrm{d}t},
   \end{eqnarray}
   where \(F(R,t)\) is the time-dependent force acting on the particle.
   Above differential equations represent an initial value problem for which one has to specify initial conditions for \(R\) and \(P\).
   The initial conditions are sampled from the Wigner distribution
 of the vibrational ground-state wave function $\psi_0(R)$, 
 \begin{eqnarray}
 W(R,P)=\frac{1}{2\pi}\int \psi_0^{*}\left(R+\frac{R'}{2}\right)\, \psi_0\left(R-\frac{R'}{2}\right)\, e^{\rmi PR'} \, \mathrm{d}R'. \label{eq:Wigner distribution}
 \end{eqnarray}
 The propagation of the trajectories is performed using the fourth-order Runge-Kutta method \cite{Runge1895,Kutta1901} using adaptive step size with on-the-fly linear interpolation of the dipole and potential-energy curves along the $R$-grid.
 Trajectories reaching \(R > \SI{100}{\au}\) within the duration of the laser pulse are considered as dissociated. For the remaining trajectories, dissociation is defined as having final total energy above the asymptotic value of the ground-state potential-energy curve. The dissociation yield is measured by the number of dissociated trajectories divided by the total number of initial trajectories. 

\section{Results and discussion}

\begin{figure}
    \centering
    \includegraphics{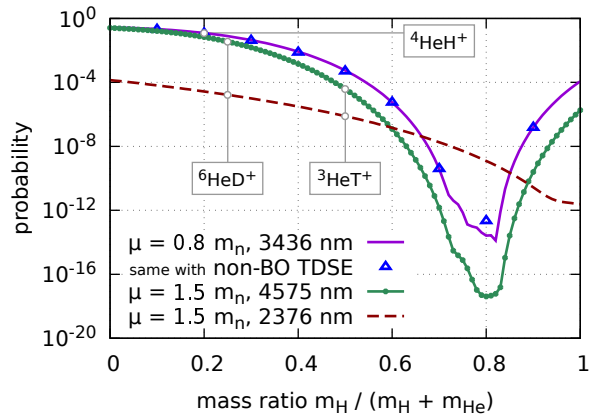}
    \caption{Dissociation probabilities in the electronic ground state 
    calculated from the electron-nuclear TDSE (blue triangles) and from the two-level nuclear TDSE (all other curves, 101 data points each).
    The laser pulse is \SI{50}{fs} long with \SI{7e13}{W/cm^2} peak intensity.
    The wavelength is chosen to closely match the $v=0 \to 1$ transition (violet line, blue triangles and small green circles) or $v = 0 \to 2$ transition (brown dashed curve). The initial state is $v=0$.
    Data points that correspond to existing isotopologues of \HeHp are marked.
    }
    \label{fig:yield_near_resonance}
\end{figure}

The dissociation yield as a function of the mass ratio $r$ is shown in \fref{fig:yield_near_resonance}.
We choose laser frequencies to closely match the resonance $v = 0 \to 1$ or $v = 0 \to 2$. (Here $v$ is the vibrational quantum number.) Despite always matching a resonant transition, the dissociation yield changes by many orders of magnitude as a function of $r$.
The agreement between the Born-Oppenheimer nuclear and non-Born-Oppenheimer electron-nuclear TDSE is very good, indicating that effects from higher electronic states (beyond the first excited state) are negligible.
There is a notable minimum around $r = 0.8$ for the $v = 0 \to 1$ resonance whereas the yield decreases monotonically with $r$ for the $v = 0 \to 2$ resonance case.

With increasing $r$, the dipole coupling $d(R)$  becomes monotonically smaller for most $R$ as can be seen in \fref{fig:dipole_coupling}.
In a very simple picture where \HeHp consists of a neutral helium atom and a proton, the dipole coupling is $d(R) = (1-r)R$.
In this case,
increasing $r$ effectively has the same effect as decreasing the amplitude of the electric field $E(t)$ in the nuclear Hamiltonian $H_1$, see equation~\eref{eq:nuclear_Hamiltonian}.
The exact value of $d(R)$ differs somewhat because the ground-state electron is not exactly located at the helium nucleus, giving rise to the “bump” in $d(R)$.

\begin{figure}
    \centering
    \includegraphics{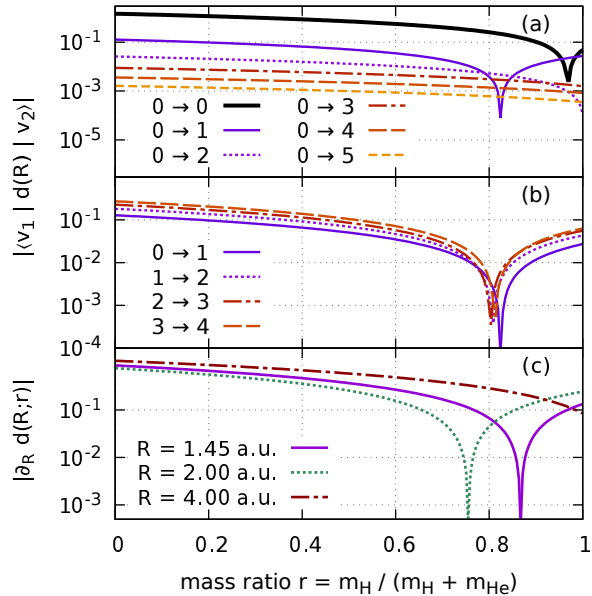}
    \caption{(a) and (b): Vibrational transition matrix elements \(|\langle v_1|d(R)|v_2 \rangle|\) for some vibrational transitions \(v_1 \to v_2\) in the electronic ground state with \(\mu = 0.8 m_n\).
    (a) Transitions from the vibrational ground state to other vibrational states.
    (b) First four transitions in a vibrational ladder-climbing
    scheme starting in the ground state \(v = 0\).
    (c) Derivative of the dipole coupling \(d(R)\) at three fixed internuclear distances. This quantity is proportional to the classical external driving force, see text.
    }
   
    \label{fig:vibrational_transition_matrix_elements}
\end{figure}

As a result, the coupling strengths (dipole matrix elements) of some vibrational transitions show distinct minima as a function of $r$, see \fref{fig:vibrational_transition_matrix_elements}.
For the series of vibrational transitions that are necessary for vibrational ladder climbing, \ie $v = 0 \to 1 \to 2$, etc., several minima close to $r = 0.8$ play together (see \fref{fig:vibrational_transition_matrix_elements}(b)) to create the structure in the dissociation yield in \fref{fig:yield_near_resonance}.
Note that due to the anharmonicity of the potential, successive transitions between higher vibrational states are not in resonance for the chosen wavelength, thus some states can be skipped
and several excitation pathways to dissociation may be utilized, some even with similar probabilities.

However, excitation to the first excited vibrational state is a gateway for all relevant dissociation pathways in the $v = 0 \to 1$ resonance case.
This gateway is blocked for a certain mass ratio, leading to substantial suppression of the dissociation yield.
Note that the dissociation yield in the $v = 0 \to 2$ resonance case in \fref{fig:yield_near_resonance} follows a similar trend as the $v = 0 \to 2$ curve in \fref{fig:vibrational_transition_matrix_elements}(a).

\begin{figure}
    \centering
    \includegraphics{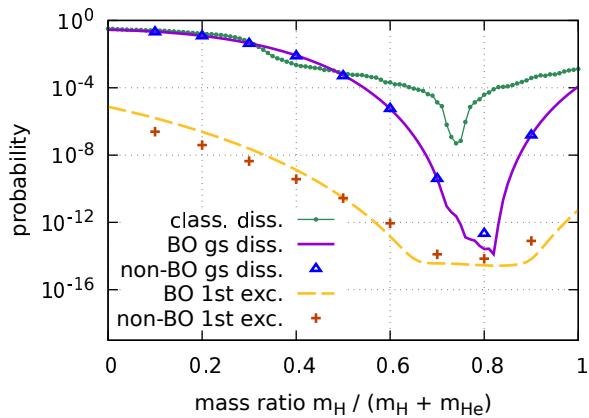}
    \caption{
    Dissociation probabilities
    for molecules with $\mu = 0.8 m_n$ in a \SI{50}{fs} laser pulse with \SI{7e13}{W/cm^2} peak intensity at \SI{3436}{nm} wavelength.
    The green circles show results from classical calculations on the electronic-ground-state potential curve. 
    The violet solid curve and the blue triangles show the ground-state dissociation yield from TDSE calculations (same as in \fref{fig:yield_near_resonance}).
    Additionally, the population of the first excited electronic state is shown (yellow dashed curve and orange crosses).
    The curves without symbols (101 data points each) are calculated by solving the two-level nuclear TDSE; the blue triangles and orange crosses are results from the electron-nuclear TDSE.
    }
    \label{fig:excitation+classical_yield}
\end{figure}

The dissociation probabilities from classical calculations are shown as green circles/line together with the corresponding quantum-mechanical results in \fref{fig:excitation+classical_yield}.
While the suppression of dissociation is not as strong, the qualitative behaviour is similar to the TDSE simulation.
The classical analogue to the quantum-mechanical explanation via dipole transition matrix elements is shown in \fref{fig:vibrational_transition_matrix_elements}(c).
The force that a classical particle on the ground-state potential-energy curve experiences from the laser field is proportional to \(d'(R)\).
As a function of the mass ratio \(r\), it is roughly linear (see \eref{eq:dipole_coupling}) and may or may not cross zero, depending on the internuclear distance.
Three examples for this behaviour are shown in \fref{fig:vibrational_transition_matrix_elements}(c).
At the equilibrium distance (\(R = \SI{1.45}{\au}\))
and on the particle’s way out to the dissociation continuum 
it can be trapped when the coupling force vanishes, giving rise to the drop in dissociation yield around \(r = \num{0.75}\).

By looking at \eref{eq:TDSE_two-level} and the definition of $d_{12}$, it could be expected that the excitation to the first excited electronic state is independent of $r$.
Instead, we notice that the yield in the first excited state (lower curve and points in \fref{fig:excitation+classical_yield}) qualitatively follows the ground-state dissociation yield (upper curve and points) and varies over many orders of magnitude.
The plateau at approximately \num{e-15} is probably caused by limited numerical accuracy and is not physically relevant.
These numerical results indicate that the main pathway to the electronically excited state is not the direct electronic excitation from the initial vibrational state but instead enhanced excitation at larger internuclear distance. Clearly, the expansion of the molecule to larger internuclear distance  happens primarily when also the dissociation yield is high.
Enhanced excitation at certain internuclear distances is already known for asymmetric molecules 
\cite{PhysRevA.76.053409}.
The need for nuclear motion as the initial step preceding electronic excitation 
has also been identified in the ionization channel of \HeHp \cite{PhysRevLett.121.073203}.

\section{Conclusions}

In this work, we have presented numerical results for the  dissociation of artificial diatomic molecular ions driven by strong laser pulses. We have found that the dissociation probability is highly dependent on the nuclear mass ratio, even when the reduced nuclear mass is kept constant. 
For sufficiently long wavelengths, the dissociation yield exhibits a distinct minimum at a certain mass ratio where the the transition dipole moments between vibrational states are strongly suppressed.
Classically, a similar suppression occurs because the laser-induced force on the nuclei is small at certain combinations of mass ratio and internuclear distance.
\manfred{In additional calculations not shown here, we have confirmed that the effect is present whenever the laser wavelength is in the vicinity of the resonance wavelength of the lowest vibrational transition and also for longer wavelengths. For shorter pulse durations such as 10$\,$fs, i.e. at larger laser bandwidths, the minimum in the mass-ratio dependent curve is broadened, but the suppression still covers many orders of magnitude.}
The effect described here is not due a variation of the reduced mass, which is another important parameter determining the nuclear motion.
The suppression of dissociation can be traced back to the dependence of the field-molecule coupling on the location of the nuclear center of mass. This dependence arises because a molecular ion is a charged system, for which the electric dipole depends on the choice of reference point.
\manfred{More generally, if we compare different isotopologues of a molecular ion (such as $^4$HeH$^+$ and $^4$HeD$^+$), both the reduced mass and the mass ratio may vary, but nonetheless the mass-ratio effect will be present and important for the understanding of the observables.}
\manfred{Unlike many previous studies that focused on near-infrared laser pulses, we have concentrated here on relatively long wavelengths in the mid-infrared range. In view of the increasing research activity related to strong mid-infrared fields \cite{Wolter2015}, we anticipate that the production of molecular ions by ionization of neutral molecules and the subsequent laser-induced photodissociation will play an important role in future studies, so our results will contribute to the interpretation of future experiments.}

\section{Acknowledgements}

We are grateful to the Deutsche Forschungsgemeinschaft for supporting this work through the Priority Programme 1840, Quantum Dynamics in Tailored Intense Fields (QUTIF). 

\section*{References}
\bibliography{refs.bib}
\end{document}